\documentclass{PoS}

\usepackage{booktabs}

\title{Photon Reconstruction for H.E.S.S. Using a Semi-Analytical Shower Model}

\ShortTitle{Photon Reconstruction for H.E.S.S. Using a Semi-Analytical Shower Model}

\author{\speaker{M. Holler}$^{a}$, A. Balzer$^{b}$, R. Chalm\'e-Calvet$^{c}$, M. de Naurois$^{a}$, D. Zaborov$^{a}$\thanks{On leave from Institute for Theoretical and Experimental Physics, B. Cheremushkinskaya 25, Moscow, 117218 Russia}~, for the H.E.S.S. collaboration\\
        ${}^{a}$Laboratoire Leprince-Ringuet -  \'Ecole Polytechnique\\
        ${}^{b}$Anton Pannekoek Institute for Astronomy\\
        ${}^{c}$LPNHE Paris\\
        E-mail: \email{holler@llr.in2p3.fr}}

\abstract{The High Energy Stereoscopic System (H.E.S.S.) is an array of five Imaging Atmospheric Cherenkov Telescopes (IACTs) designed to detect cosmogenic gamma-rays with very high energies. Originally consisting of just four identical IACTs (CT1-4) with an effective mirror diameter of $12\,\mathrm{m}$ each, it was expanded with a fifth IACT (CT5) with a mirror diameter of $28\,\mathrm{m}$ in 2012. Being the largest IACT worldwide, CT5 allows to lower the energy threshold of H.E.S.S., making the array sensitive at energies where space-based detectors run out of statistics. Events can be analysed either monoscopically (i.e. using only information of CT5) or stereoscopically (requiring at least two triggered telescopes per event). To achieve a good performance, a sophisticated event reconstruction and analysis framework is indispensable. This is particularly important for H.E.S.S. since it is now the first IACT array that consists of different telescope types. An advanced reconstruction method is based on a semi-analytical model of electromagnetic particle showers in the atmosphere (``model analysis''). The properties of the primary particle are reconstructed by comparing the image recorded by each triggered telescope with the Cherenkov emission from the shower model using a log-likelihood maximisation. Due to its performance, this method has become one of the standard analysis techniques applied to CT1-4 data. Now it has been modified for use with the five-telescope array. We present the adapted model analysis and its performance in both monoscopic and stereoscopic analysis mode.}

\FullConference{The 34th International Cosmic Ray Conference,\\
		30 July- 6 August, 2015\\
		The Hague, The Netherlands}

\begin{document}

\newcommand{\hess}{H.E.S.S.}

\section{Introduction}

Within the last decade, the field of ground-based gamma-ray astronomy has evolved from an experimental niche with only few detected sources to a mature branch of astrophysics. The High Energy Stereoscopic System (\hess ) has been playing a key role in this transition process. It was inaugurated in 2004 with four $12\,\mathrm{m}$-size Imaging Atmospheric Cherenkov Telescopes (IACTs) located in Namibia, marking the beginning of \hess\ I. The greater part of the currently known very high energy (VHE, $E~\gtrsim~100\,\mathrm{GeV}$) gamma-ray sources were detected with these four IACTs (named CT1-4). In 2012, a fifth telescope (CT5) was added in the centre of the array, initiating \hess\ II. With its large mirror with a size of $24\,\mathrm{m} \times 32\,\mathrm{m}$, it is not only able to detect electromagnetic showers which are too faint to be seen by CT1-4, but also makes \hess\ the only IACT array that consists of different telescope types and sizes. A capable event reconstruction and analysis framework is hence necessary to live up to the success of \hess\ I.

In this contribution, we present the successful adaptation of the model analysis \cite{2009_deNaurois} to the full five-telescope \hess\ array. The focus is set on the illustration and comparison of different reconstruction and analysis modes. After describing the revised model analysis in Section~\ref{model}, three different analysis modes of \hess\ II data are tested on the Crab Nebula in Section~\ref{crab}. After that, the performance is evaluated in Section~\ref{performance}.

\section{The Model Analysis}
\label{model}

In 2009, \cite{2009_deNaurois} introduced an advanced event reconstruction method for IACTs\footnote{This article is the baseline reference of this section.}. Its foundation is a semi-analytical model that describes the average development of electromagnetic particle showers in the Earth's atmosphere as a function of the primary energy of a gamma-ray ($E$) and its depth of first interaction ($T$). Building up on this model, the amount of Cherenkov emission of a shower in the camera of an IACT is analytically calculated. Templates are generated and stored for different zenith angles, impact distances\footnote{The impact distance corresponds to the distance of the projected shower axis to the telescope.} $R$, $E$, and $T$. For the event reconstruction, the intensities as measured by the camera of an IACT are compared to the shower templates by calculating a likelihood for each pixel:
\begin{equation}
\label{eq_jpt}
P(s|\mu,\sigma_p,\sigma_\gamma,\sigma_c) = \sum_n \frac{\mu^n e^{-\mu}}{n!\sqrt{2\pi (\sigma_p^2 + n \sigma_\gamma^2 + n^2 \sigma_c^2)}} \exp\left(  - \frac{(s - n)^2}{2 (\sigma_p^2 + n \sigma_\gamma^2 + n^2 \sigma_c^2)} \right)~\textrm{,}
\end{equation}
where $s$ denotes the measured signal intensity and $\mu$ the expectation value from the model templates, the latter being interpolated between the grid points of the templates. $\sigma_p$ corresponds to the pedestal width, $\sigma_{\gamma}$ to the width of the single photoelectron peak, and $\sigma_c$ accounts for possible calibration uncertainties. To compare measured and modelled intensities for a set of pixels (e.g. all pixels of a telescope or the whole array), a log-likelihood function is defined as
\begin{equation}
\ln {\it L}_{\mathrm{set}} = \sum_{\mathrm{pixel}\ i}  \ln {\it L}_i = \sum_{\mathrm{pixel}\ i} -2 \times \ln P(s_i|\mu_i,\sigma_p,\sigma_\gamma,\sigma_c)~\mathrm{.} 
\label{likelihood}
\end{equation}
To reconstruct the parameters of a primary particle (assuming that it is a gamma-ray), the log-likelihood is optimised using a Levenberg-Marquardt algorithm (see \cite{1944_Levenberg} and \cite{1963_Marquardt}). The result of the optimisation is an estimate of the arrival direction $Dir$, $E$, $T$, and $R$. Additionally the uncertainty of each parameter is assessed by calculating the second derivatives of $\ln {\it L}$. As an example, $\Delta Dir~=~1 / \sqrt{\partial^2 \ln L / \partial {Dir}^2}$ provides an event-wise estimate on how well the direction was reconstructed.

\begin{figure}
\center
\includegraphics[width = 0.7 \textwidth]{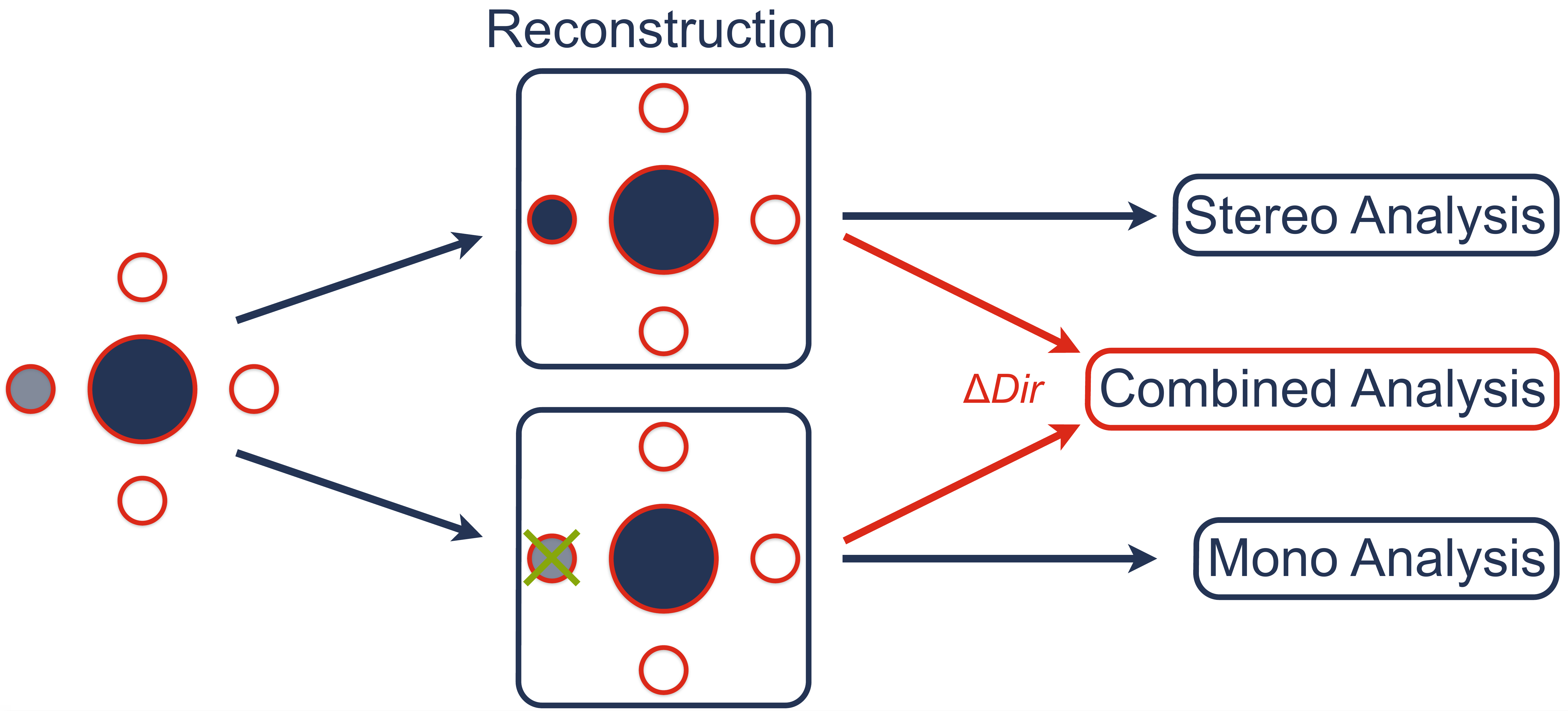}
\caption{Sketch that illustrates the different reconstruction and analysis modes (see text for further explanation).}
\label{sketch_modes}
\end{figure}

During \hess\ phase I, at least two telescopes of the array were required to be triggered by a shower for the readout of an event, and thus no single-telescope reconstruction was performed. Since the inauguration of CT5, \hess\ reads out both stereoscopic events from CT1-5 and monoscopic events from CT5. The reconstruction of monoscopic events from such a large telescope allows to significantly lower the energy threshold. Fig.~\ref{sketch_modes} shows the different reconstruction and analysis modes for the \hess\ II model analysis. The stereoscopic events where CT5 triggered are reconstructed both monoscopically and stereoscopically. In addition to the \textit{Mono} and \textit{Stereo} analysis modes, we introduce a \textit{Combined} analysis mode which makes use of both reconstruction approaches. In case an event was fitted both monoscopically and stereoscopically, the respective $\Delta Dir$ values are used to decide which result to use for the \textit{Combined} mode.

To separate cosmic gamma-rays from the excessive background which is mainly caused by particle showers of charged cosmic rays, the \textit{Shower Goodness} compares the optimised log-likelihood of the shower region with its mean expectation value (see \cite{2009_deNaurois} for the calculation of the latter):
\begin{equation}
G_{\mathrm{SG}} = \frac{\sum_{i} \left[ \ln L \left( s_i | \mu_i \right) - \left< \ln L \right>|_{\mu_i} \right]}{\sqrt{2 \cdot \textrm{NdF}}}~\textrm{,}
\end{equation}
where NdF denotes the number of degrees of freedom and $\ln L$ is calculated as defined in Eq.~\ref{likelihood}. $G_{\mathrm{SG}}$ is designed to behave like a normal variable. For stereoscopic reconstruction, the \textit{Shower Goodness} values of the individual telescopes are averaged to obtain one overall \textit{Shower Goodness}.

To reject events that are compatible with night-sky background (NSB) noise, the \textit{NSB Goodness} is defined as
\begin{equation}
G_{\mathrm{NSB}} = \frac{\sum_{i} \left[ \ln L \left( s_i | 0 \right) - \left< \ln L \right>|_{0} \right]}{\sqrt{2 \cdot \textrm{NdF}}}~\textrm{.}
\end{equation}
The main purpose of this variable is to adjust the energy threshold of the analysis, which is needed to limit systematic uncertainties. 

\begin{table}
\centering
\begin{tabular}{lccc}\toprule
 & Mono & Stereo & Combined  \\ \midrule
$G_{\mathrm{SG}}$ & $[-4,0.6]$ & $[-4,0.9]$ & $[-4,0.9]$ \\
$G_{\mathrm{NSB}}$ & $> 32$ & $> 28$ & $> 32$ \\
$T$ & $[-1.1,1.3]$ & $[-1.1,3.4]$ & $[-1.1,3.4]$ \\
$\Delta Dir$ & $< 0.3^{\circ}$ & $< 0.2^{\circ}$ & $< 0.3^{\circ}$ \\
$\vartheta^{2} (\mathrm{deg}^2)$ & $< 0.015$ & $< 0.006$ & $< 0.015$ \\
\bottomrule
\end{tabular}
\caption{Acceptance ranges of the respective \textit{Standard} cut configurations for the three different analysis modes. The respective $\vartheta^2$ cuts were optimised for point-like sources.}
\label{tab_cuts}
\end{table}
For each analysis mode (Mono, Stereo, and Combined) a \textit{Standard} cut configuration was defined (see Table~\ref{tab_cuts}). The event selection cuts, except for the cut on $G_{\mathrm{NSB}}$, were optimised for maximum significance on a weak point source with a power-law spectrum observed at a zenith angle of $18^{\circ}$. The photon index was set to $3.0$ for optimising the \textit{Mono Standard} cuts, and $2.5$ for \textit{Stereo Standard}. Additional cut configurations were defined to maximise the discovery potential for sources with other spectral indices and/or cutoffs. Systematic uncertainties related to the event selection cuts were included in consideration, with a minor effect on the final choice of the analysis cuts. The cuts for the \textit{Combined} mode have not been optimised yet but correspond to a compromise between the ones of the \textit{Mono} and \textit{Stereo} modes.

\section{Application to the Crab Nebula}
\label{crab}

To test the three \hess\ II model analysis modes, $7.47\,\textrm{h}$ of good-quality observations of the Crab Nebula are used. After dead-time correction, this corresponds to $7.23\,\textrm{h}$ (\textit{Mono} and \textit{Combined} mode) and $6.79\,\textrm{h}$ (\textit{Stereo} mode), respectively. The data were taken in October and November 2014 at zenith angles between $45^{\circ}$ and $55^{\circ}$. The source was observed in wobble mode with offsets of $\pm 0.5^{\circ}$ in both right ascension and declination. 

\begin{figure}
\includegraphics[width = \textwidth]{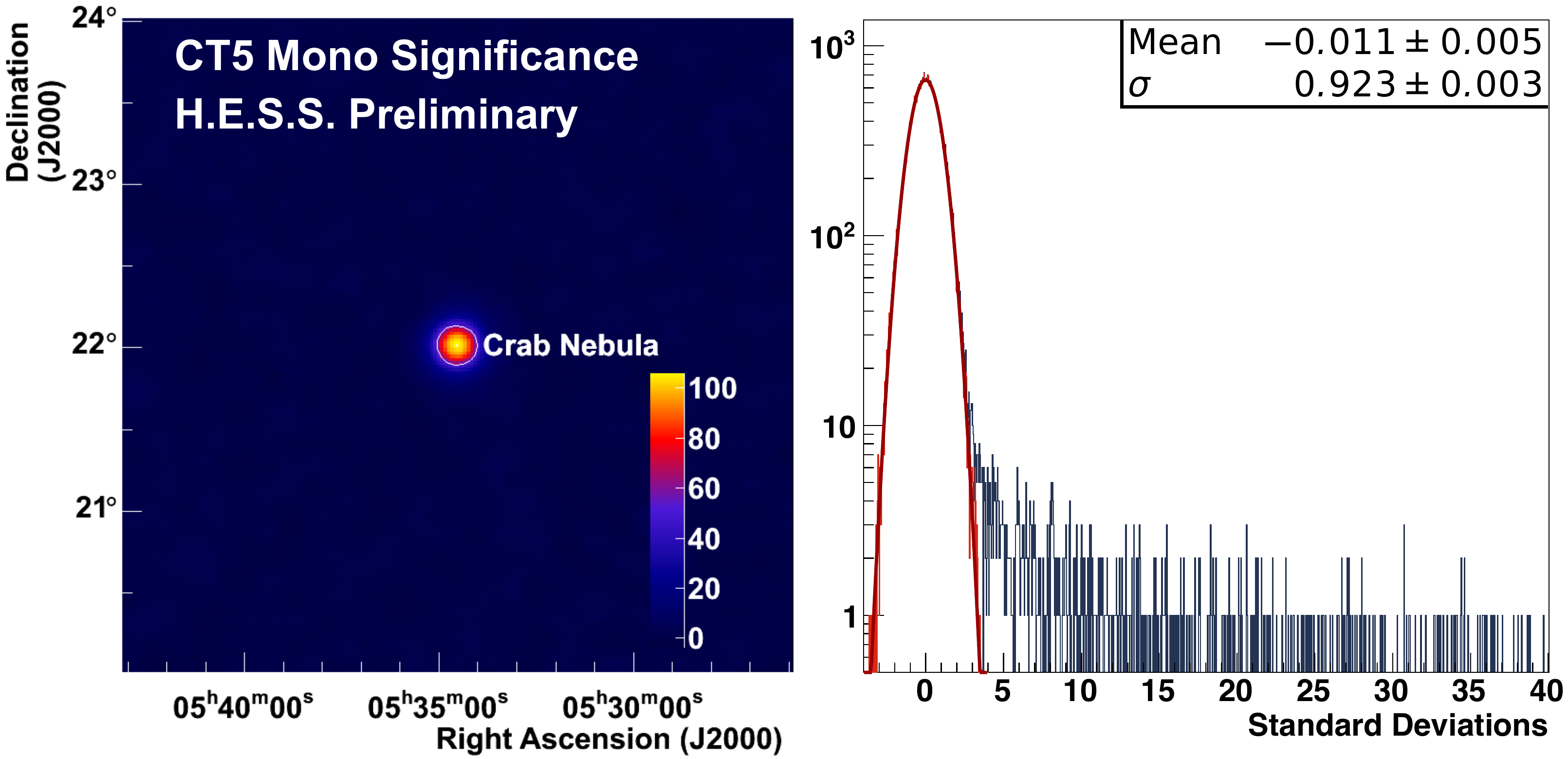}
\caption{\textit{Left:} significance map of the region around the Crab Nebula, obtained using the monoscopic analysis mode. \textit{Right:} distribution of significances for the whole map (dark blue) and the part outside the target exclusion region (red histogram). The latter was fitted with a Gaussian (solid red line; see fit results on the top right). Only significances up to $40\sigma$ are shown for visibility reasons.}
\label{Map_Mono}
\end{figure}
The significance sky map as obtained with the ring background method \cite{2007_Berge_Background} in \textit{Mono} mode is shown in the \textit{left panel} of Fig.~\ref{Map_Mono}. It shows a radially symmetric point-like source (detailed comparisons of the simulated and measured angular resolution still have to be performed though) that is centered on the position of the Crab Nebula. The map is very well normalised (see \textit{right panel} of Fig.~\ref{Map_Mono}), indicating that background subtraction errors are well under control.

\begin{figure}
\center
\includegraphics[width = 0.5\textwidth]{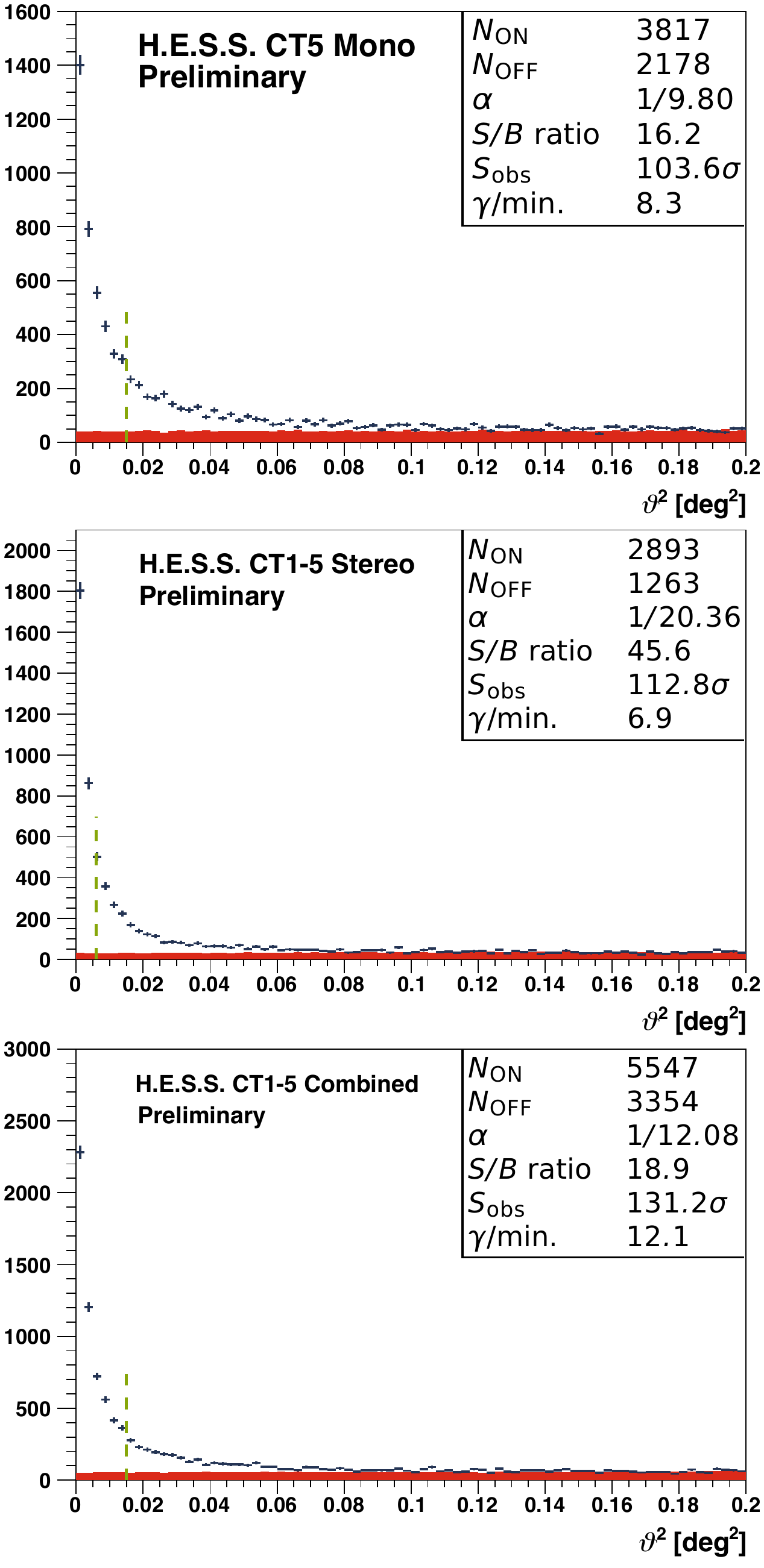}
\caption{Angular distribution of $\gamma$-like events from the Crab Nebula for the monoscopic (\textit{top}), stereoscopic (\textit{middle}), and combined (\textit{bottom}) analysis mode, respectively. In each panel, the distribution around the source is shown in dark blue, and the one from the background regions (scaled with $\alpha$) in red. The $\vartheta^{2}$ cut is illustrated with a dashed green line, and the statistics are given on the top right.}
\label{Theta2}
\end{figure}
Fig.~\ref{Theta2} shows the angular distributions and event statistics of gamma-like events for all three analysis modes. For each mode, the source and background distributions are in good agreement at large $\vartheta^2$ values. The \textit{Combined} mode yields the highest excess rate and significance whereas the \textit{Stereo} mode performs best in terms of $S/B$ ratio.

\section{Performance}
\label{performance}

In the following, the performance of the three modes for low-zenith observations is evaluated using Monte-Carlo (MC) simulations. The simulations correspond to a point source at wobble offset $0.5^{\circ}$, zenith angle $18^{\circ}$, and azimuth angle $180^{\circ}$ (i.e. southward pointing). The results of this study should not be directly compared to the ones of the previous section since the Crab Nebula can just be observed at zenith angles above $45^{\circ}$ with \hess .

\begin{figure}
\center
\includegraphics[width = 0.8\textwidth]{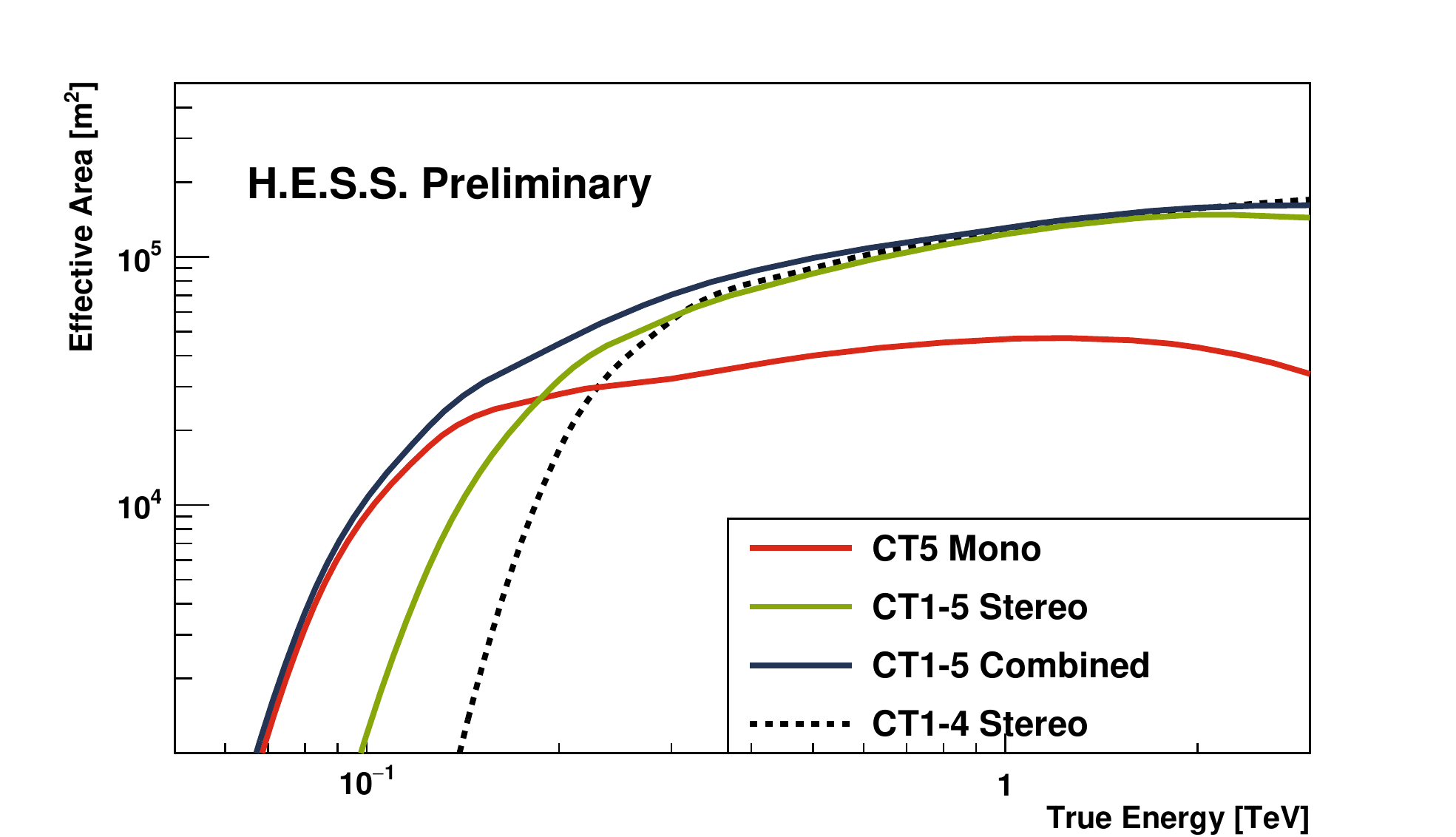}
\caption{Effective area (after selection cuts) of the model analysis for the different analysis modes presented here. The curves were computed using simulations with a zenith angle of $18^{\circ}$ and an azimuth angle of $180^{\circ}$.}
\label{Aeff}
\end{figure}
The post-cut effective area of the \hess\ II model analysis for the respective \textit{Standard} cuts of the three analysis modes is shown in Fig.~\ref{Aeff}. The effective area of CT1-4 \textit{Stereo} (\hess\ I) is drawn for comparison. Whilst being almost identical at $E \gtrsim 300\,\mathrm{GeV}$, the curves of CT1-4 \textit{Stereo}, CT1-5 \textit{Stereo}, and CT1-5 \textit{Combined} diverge below. By construction, the \textit{Combined} mode joins the low energy threshold of the \textit{Mono} analysis with the high detection area of the \textit{Stereo} mode at higher energies. It thus provides the best energy coverage.

To assess the overall performance of the revised model analysis, the differential sensitivity was calculated. Here it is defined as the minimum source strength to obtain a $5\sigma$ detection after $50\,\mathrm{h}$ of observation time at a given energy range. For the calculation of the significance, the simplified method $N_{\gamma}/\sqrt{N_{\textrm{bkg}}}$ was used. Background events were taken from actual observations with zenith angles between $12^{\circ}$ and $22^{\circ}$ to closely match the simulations.
\begin{figure}
\includegraphics[width = \textwidth]{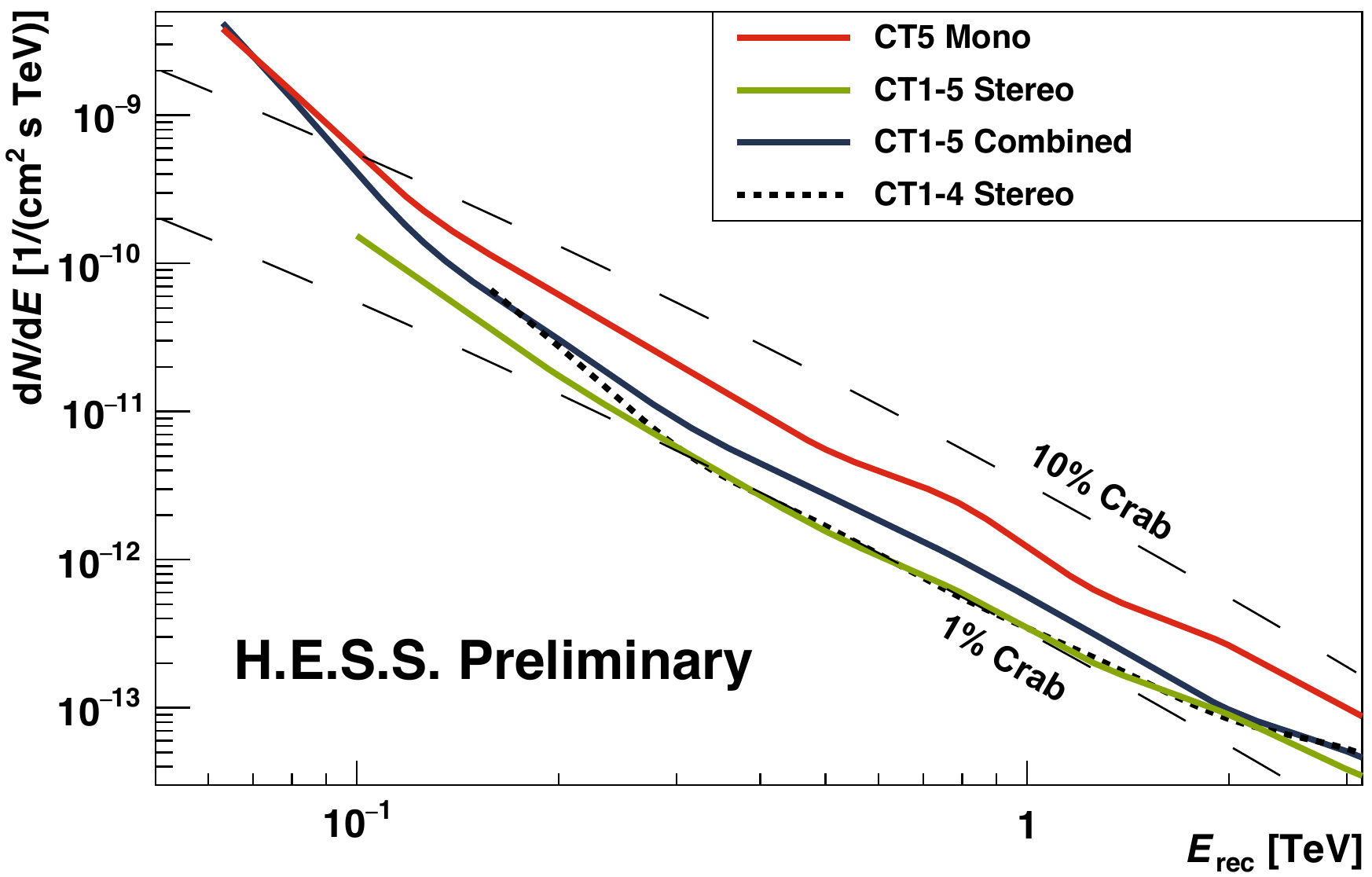}
\caption{Differential sensitivity of the model analysis for the three analysis modes of H.E.S.S. II data, as well as for \hess\ I for comparison. All curves correspond to point-like sources. The reference spectrum of the Crab Nebula used to draw the dashed black lines was taken from \cite{2015_MAGIC_Spectrum}.}
\label{Diff_Sens}
\end{figure}
The differential sensitivities for the different modes are shown in Fig.~\ref{Diff_Sens}. Each of the curves was calculated using five bins per decade. A minimum of $10$ excess counts as well as a minimum $S/B$ ratio of $0.05$ was required for each bin. As already expected from the effective area curves, the \textit{Combined} mode always performs at least as good as the \textit{Mono} one. At medium and high energies it is however currently not as sensitive as the \textit{Stereo} mode, which can partly be attributed to the fact that its cuts are not yet optimised. The larger $\vartheta^2$ cut for instance directly translates into an increased background level.

\section{Conclusions}
\label{conclusions}

We presented the successful adaptation of the advanced model analysis \cite{2009_deNaurois} to \hess\ phase II. In addition to the \textit{Mono} and \textit{Stereo} analysis modes, we introduced the first combined analysis of both monoscopic and stereoscopic events for an IACT instrument. This method provides the best energy coverage and makes use of all the information recorded by the five-telescope \hess\ array.

The analysis modes were successfully tested on the Crab Nebula, and all provide trustworthy results on this standard source. For complementation, the effective area and differential sensitivity were evaluated for sources with low zenith angles using MC simulations and background events from actual data. Though some optimisation still has to be carried out for the \textit{Combined} mode, all results shown here prove that the updated model analysis for \hess\ II continues to be a very sensitive and advanced method in the field of ground-based gamma-ray astronomy. 

\section{Acknowledgements}

The support of the Namibian authorities and of the University of Namibia in facilitating the construction and operation of H.E.S.S. is gratefully acknowledged, as is the support by the German Ministry for Education and Research (BMBF), the Max Planck Society, the German Research Foundation (DFG), the French Ministry for Research, the CNRS-IN2P3 and the Astroparticle Interdisciplinary Programme of the CNRS, the U.K. Science and Technology Facilities Council (STFC), the IPNP of the Charles University, the Czech Science Foundation, the Polish Ministry of Science and Higher Education, the South African Department of Science and Technology and National Research Foundation, and by the University of Namibia. We appreciate the excellent work of the technical support staff in Berlin, Durham, Hamburg, Heidelberg, Palaiseau, Paris, Saclay, and in Namibia in the construction and operation of the equipment.

\bibliographystyle{JHEP}
\bibliography{model_proceeding}

\end{document}